# ARMA Time-Series Modeling with Graphical Models

**Bo Thiesson**
Microsoft Research
thiesson@microsoft.com

**David Maxwell Chickering**
Microsoft Research
dmax@microsoft.com

**David Heckerman**
Microsoft Research
heckerma@microsoft.com

**Christopher Meek**
Microsoft Research
meek@microsoft.com

## Abstract

We express the classic ARMA time-series model as a directed graphical model. In doing so, we find that the deterministic relationships in the model make it effectively impossible to use the EM algorithm for learning model parameters. To remedy this problem, we replace the deterministic relationships with Gaussian distributions having a small variance, yielding the stochastic ARMA ($\sigma$ARMA) model. This modification allows us to use the EM algorithm to learn parameters and to forecast, even in situations where some data is missing. This modification, in conjunction with the graphical-model approach, also allows us to include cross predictors in situations where there are multiple time series and/or additional non-temporal covariates. More surprising, experiments suggest that the move to stochastic ARMA yields improved accuracy through better smoothing. We demonstrate improvements afforded by cross prediction and better smoothing on real data.

## 1 Introduction

Graphical models have been used to represent time-series models for almost two decades (e.g., Dean and Kanazawa, 1988; Cooper, Horvitz, and Heckerman, 1988). The benefits of such representation include the ability to apply standard graphical-model-inference algorithms for parameter estimation (including estimation in the presence of missing data), for prediction, and for cross prediction—the use of one time series to help predict another time series (e.g., Ghahramani, 1998; Meek, Chickering, and Heckerman, 2002; Bach and Jordan, 2004).

In this paper, we express the classic autoregressive moving average (ARMA) time-series model (e.g., Box, Jenkins, and Reinsel, 1994) as a graphical model to achieve similar benefits. As we shall see, the ARMA model includes deterministic relationships, making it effectively impossible to estimate model parameters via the Expectation–Maximization (EM) algorithm. Consequently, we introduce a variation of the ARMA model for which the EM algorithm can be applied. Our modification, called stochastic ARMA ($\sigma$ARMA), replaces the deterministic component of the ARMA model with a Gaussian distribution having a small variance. To our surprise, this addition not only has the desired effect of making the EM algorithm effective for parameter estimation, but our experiments also suggest that it provides a smoothing technique that produces more accurate estimates.

We have chosen to focus on ARMA time-series models in this paper, but our paper applies immediately to autoregressive *integrated* moving average (ARIMA) time-series models as well. In particular, an ARIMA model for a time series is simply an ARMA model for a preprocessed version of that same time series. This preprocessing consists of $d$ consecutive differencing transformations, where each transformation replaces the observations with the differences between successive observations. For example, when $d = 0$ an ARIMA model is a regular ARMA model, when $d = 1$ an ARIMA model is an ARMA model of the differences, and when $d = 2$ an ARIMA model is an ARMA model of the differences of the differences. In practice, $d \leq 2$ is almost always sufficient for good results (Box, Jenkins, and Reinsel, 1994).

In Section 2, we review the ARMA model and introduce our stochastic variant. We also extend the model to allow for cross prediction, yielding $\sigma$ARMA$^{xp}$, a model more general than previous extensions to cross-prediction ARMA (e.g., the vector ARMA model; Reinsel, 2003). In Section 3, we describe how the EM algorithm can be applied to $\sigma$ARMA$^{xp}$ for parameter estimation. Our approach is, to our knowledge, the first EM-based alternative to parameter estimation in the ARMA model class. In Section 4, we describe how to predict future observations—that is, forecast—using $\sigma\text{ARMA}^{xp}$. In Sections 5 and 6, we demonstrate the utility of our extensions in an evaluation using two real data collections. We show that (1) the stochastic variant of ARMA produces more accurate predictions than those of standard ARMA, (2) estimation and prediction in the face of missing data using our approach yields better forecasts than by a heuristic approach in which missing data are filled in by interpolation, and (3) the inclusion of cross predictions can lead to more accurate forecasting.

## 2 Time-Series Models

In this section, we review the well-known class of autoregressive-moving average (ARMA) time series models and define two closely related stochastic variations, the $\sigma\text{ARMA}$ and the $\sigma\text{ARMA}^*$ classes. We also define a generalization of these two stochastic variations, called $\sigma\text{ARMA}^{xp}$ and $\sigma\text{ARMA}^{*xp}$, which additionally allow for selective cross-predictors from related time series.

We begin by introducing notation and nomenclature. We denote a temporal sequence of observation variables by $Y = (Y_1, Y_2, \ldots, Y_T)$. Time-series data is a sequence of values for these variables denoted by $y = (y_1, y_2, \ldots, y_T)$. We suppose that these observations are obtained at discrete, equispaced intervals of time. In this paper we will consider incomplete observation sequences in the sense that some of the observation variables may have missing observations. For notational convenience, we will represent such a sequence of observations as a complete sequence, and it will be clear from context that this sequence has missing observations.

In the time-series models that we consider in this paper, we associate a latent "white noise" variable with each observable variable. These latent variables are denoted $E = (E_1, E_2, \ldots, E_T)$.

Some of the models will contain cross-predictor sequences. A cross-predictor sequence is a sequence of observation variables from a related time series, which is used in the predictive model for the time series under consideration. For each cross-predictor sequence in a model, a cross-predictor variable is associated with each observable variable. For instance, $Y' = (Y'_1, Y'_2, \ldots, Y'_{T'})$ and $Y'' = (Y''_1, Y''_2, \ldots, Y''_{T''})$ may be related time series sequences where $Y'_{t-1}$, $Y'_{t-12}$ and $Y''_{t-1}$ are cross-predictor variables for $Y_t$. Let $C_t$ denote a vector of cross-predictor variables for $Y_t$. The set of cross-predictor vectors for all variables $Y$ is denoted $C = (C_1, C_2, \ldots, C_T)$.

Our stochastic time-series models handle incomplete time-series sequences in the sense that some values in the sequence are missing. Hence, in real-world situations where the length of multiple cross-predicting time series do not match, we have two possibilities. We can introduce observation variables with missing values, or we can shorten a sequence of observation variables, as necessary. In the following we will assume that $Y$, $E$, and $C$ are all of the same length.

For any sequence, say $Y$, we denote the sub-sequence consisting of the $i$'th through the $j$'th element by $Y_i^j = (Y_i, Y_{i+1}, \ldots, Y_j)$, $i < j$.

### 2.1 ARMA Models

In slightly different notation than usual (see, e.g., Box, Jenkins, and Reinsel, 1994 or Ansley, 1979) the ARMA$(p, q)$ time series model is defined as the deterministic relation

$$Y_t = \zeta + \sum_{j=0}^{q} \beta_j E_{t-j} + \sum_{i=1}^{p} \alpha_i Y_{t-i} \quad (1)$$

where $\zeta$ is the intercept, $\sum_{i=1}^{p} \alpha_i Y_{t-i}$ is the autoregressive (AR) part, $\sum_{j=0}^{q} \beta_j E_{t-j}$ is the moving average (MA) part with $\beta_0$ fixed as 1, and $E_t \sim \mathcal{N}(0, \gamma)$ is "white noise" with $E_t$ mutually independent for all $t$. The construction of this model therefore involves estimation of the free parameters $\zeta$, $(\alpha_1, \ldots, \alpha_p)$, $(\beta_1, \ldots, \beta_q)$, and $\gamma$.

For a constructed model, the one step-ahead forecast $\hat{Y}_t$ given the past can be computed as

$$\hat{Y}_t = \zeta + \sum_{j=1}^{q} \beta_j E_{t-j} + \sum_{i=1}^{p} \alpha_i Y_{t-i} \quad (2)$$

where we exploit that at any time $t$, the error in the ARMA model can be determined as the difference between the actual observed value and the one step-ahead forecast

$$E_t = Y_t - \hat{Y}_t \quad (3)$$

The variance for this forecast is $\gamma$.

An ARMA model can be represented by a directed graphical model (or Bayes net) with both stochastic and deterministic nodes. A node is deterministic if the value of the variable represented by that node is a deterministic function of the values for variables represented by nodes pointing to that node in the graphical representation. From the definition of the ARMA models, we see that the observable variables (the $Y$'s) are represented by deterministic nodes and the error variables (the $E$'s) are represented by stochastic nodes. The relations between variables are defined by (1) and accordingly, $Y_{t-p}, \ldots, Y_{t-1}$ and $E_{t-q}, \ldots, E_t$ all point to $Y_t$. In this paper we are interested in the *conditional likelihood models*, where we condition on the first $R = max(p,q)$ variables. Relations between variables for $t \leq R$ can therefore be ignored. The graphical representation for an ARMA(2,2) model is shown in Figure 1. It should be noted that if we artificially extend the time series back in time for R (unobserved) time steps, this model represents what is known in the literature as the *exact likelihood model*. There are alternative methods for dealing with the beginning of a time series (see, e.g., Box, Jenkins, and Reinsel, 1994).

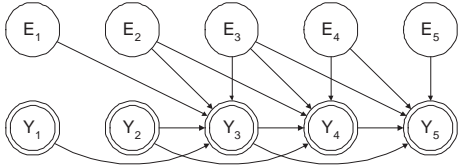

Figure 1: ARMA(2,2) model for time series with five observations. Stochastic nodes are shown as single-circles and deterministic nodes as double-circles.

## 2.2 $\sigma$ARMA Models

The EM algorithm is a standard method used to learn parameters in a graphical model. As argued in the Introduction and as we will see in Section 3.3, the EM algorithm cannot be applied to estimate the parameters in an ARMA model. In this section we define the $\sigma$ARMA class of models for which the EM algorithm can be applied. A $\sigma$ARMA model is identical to an ARMA model except that the deterministic relation in Equation 1 is replaced by a conditional Normal distribution (with variance $\sigma$, thereby the name). More specifically, for the observable sequence of variables we have $Y_t | E_{t-q}^t, Y_{t-p}^{t-1} \sim \mathcal{N}(\mu, \sigma)$, where the functional expression for the mean $\mu$ and the variance $\sigma$ are shared across the observation variables. The variance is fixed at a given (small) value to be specified by the user. In Section 4 we will see that $\sigma$ takes the role as a minimum allowed variance for the one step-ahead forecast. The mean is related to the conditional variables as follows

$$\mu = \zeta + \sum_{j=0}^{q} \beta_j E_{t-j} + \sum_{i=1}^{p} \alpha_i Y_{t-i} \quad (4)$$

We see from this representation that ARMA is the limit of $\sigma$ARMA as $\sigma \to 0$.

By fixing $\beta_0 = 1$ in the ARMA model class, the variance $\gamma$ of $E_t$ has the semantic of being the variance for the one-step forecast. As we will see in Section 4, $\gamma$ does not carry the same semantic for a $\sigma$ARMA model. Without this semantic for $\gamma$, it seems natural to extend the $\sigma$ARMA model class by additionally letting $\beta_0$ vary freely. We will denote this class of models as $\sigma$ARMA$^*$.

A $\sigma$ARMA (or $\sigma$ARMA$^*$) model has the same graphical representation as the similar ARMA model, except that deterministic nodes are now stochastic.

## 2.3 $\sigma$ARMA$^{xp}$ Models

The $\sigma$ARMA$^{xp}$ class of models includes the following generalizations of the $\sigma$ARMA class: (1) a model may define multiple time series - with different $p$'s and $q$'s in different time series and (2) a time series is allowed additional dependencies on observations from related time series, called cross-predictors (the 'xp' in the name).

Consider the part of a $\sigma$ARMA$^{xp}$ model which describes a particular time series in the set of time series defined by the model. The representation of this time series is similar to an $\sigma$ARMA model, except that $Y_t$ additionally depends on the vector of cross predictors $C_t$. Let $\eta$ be a vector of regression coefficients associated with the cross predictors. In this case $Y_t | E_{t-q}^t, Y_{t-p}^{t-1}, C_t \sim \mathcal{N}(\mu, \sigma)$ with the mean

$$\mu = \zeta + \sum_{j=0}^{q} \beta_j E_{t-j} + \sum_{i=1}^{p} \alpha_i Y_{t-i} + \eta C_t \quad (5)$$

Similar to the $\sigma$ARMA models, we define two variations. In the $\sigma$ARMA$^{xp}$ class, $\beta_0$ is fixed as 1, and in the $\sigma$ARMA$^{*xp}$, $\beta_0$ vary freely.

As mentioned above, a full $\sigma$ARMA$^{xp}$ model contains multiple time series. The time series are related through the cross predictors only, and hence the unobserved error variables are independent across time series in the model and may have different shared variances for different time series.

Besides the stochastic nature of the $\sigma$ARMA$^{xp}$ models, these models are different from vector ARMA models (see, e.g., Reinsel, 2003) in that (1) different time series in a model may have different numbers of AR and MA regressors, and (2) cross predictors between time series are not defined by the ARMA structure, but are chosen in a selective way for each time series in the model.

The graphical representation of an $\sigma$ARMA$^{xp}$ (or $\sigma$ARMA$^{*xp}$) model is shown in Figure 2. The model represents two time series. The first time series (for observations $Y_t$) correspond to an $\sigma$ARMA(2,2) model with no cross-predictors and the second time series (for observations $Y_t'$) correspond to an $\sigma$ARMA(1,1) model with a single cross-predictor, $Y_{t-1}$.

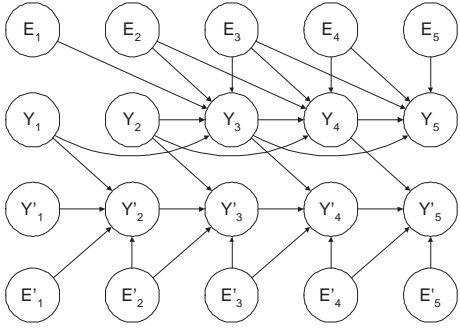

Figure 2: $\sigma\text{ARMA}^{xp}$ for two time series each with five observations.

## 3 $\sigma\text{ARMA}^{xp}$ Estimation

We seek the maximum likelihood estimate for the conditional log-likelihood of all time series in the $\sigma\text{ARMA}^{xp}$ model, where we condition on the first $R = max(p,q)$ time steps for each time series in the model. Notice that $R$ may be different for different time series, as is the case for the example in Figure 2.

This log-likelihood is, however, difficult to maximize directly due to incomplete data in the statistical model. First of all, the latent error variables renders all parameters within a given time series dependent. Secondly, some of the observable variables may have missing observations in which case cross predictors may cause parameters in different time series to become dependent.

In contrast, given (imaginary) complete data for all variables in the model, the parameters associated with each time series are independent and can therefore be estimated independently. Let $\beta = (\beta_1, \ldots, \beta_q)$ and $\alpha = (\alpha_1, \ldots, \alpha_p)$. We denote the parameters in the model by $\theta = (\theta_E, \theta_Y)$, where $\theta_E = (\gamma)$ and $\theta_Y = (\zeta, \beta, \alpha, \eta, \sigma)$ for $\sigma\text{ARMA}^{xp}$ and $\theta_Y = (\zeta, \beta_0, \beta, \alpha, \eta, \sigma)$ for $\sigma\text{ARMA}^{*xp}$. Conditioning on the first $R$ time steps, the complete data (conditional) log-likelihood for a particular time series factorizes as follows

$$
\begin{aligned}
l(\theta) &= \log p(Y_{R+1}^T, E_{R+1}^T | Y_1^R, E_1^R, C, \theta) \\
&= \sum_{t=R+1}^T \log p(Y_t | E_t, Y_{t-p}^{t-1}, E_{t-q}^{t-1}, C_t, \theta_Y) \\
&\quad + \sum_{t=R+1}^T \log p(E_t | \theta_E) \quad (6)
\end{aligned}
$$

Due to the above factorization, the complete data log-likelihood is easier to maximize. Hence, assuming that data is missing at random, the EM algorithm can be used to solve the maximization problem.

Roughly speaking, the EM algorithm converts the ML estimation problem into an iterative sequence of "pseudo-estimations" with respect to the statistical model for complete data. Let $\Theta$ denote the parameterizations $\theta$ for all the individual time series in the model and let $\Theta^n$ denote the current value of $\Theta$ after $n$ iterations. Each iteration of the EM algorithm involves two steps.

E-step: For each time series, construct the conditional expectation for the complete data log-likelihood given the current parameterization and observed data for all time series in the model

$$Q(\theta|\Theta^n) = \mathbb{E}_{\Theta^n}[l(\theta)].$$

M-step: For each time series, choose $\theta^{n+1}$ as the value that maximizes $Q(\theta|\Theta^n)$.

In this case where the statistical model is a subfamily of an exponential family, the EM algorithm becomes an alternation between an E-step that computes expected sufficient statistics for the statistical model and an M-step that re-estimates the parameters of the model by treating the expected sufficient statistics as if they were actual sufficient statistics (Dempster, Laird, and Rubin 1977).

### 3.1 $\sigma\text{ARMA}^{xp}$ (fixed $\beta_0$)

**E-step**

By the factorization (6), the expected complete data log-likelihood becomes

$$
Q(\theta|\Theta^n) = \sum_{t=R+1}^T \mathbb{E}_{\Theta^n}\left[\log p(Y_t|E_t, Y_{t-p}^{t-1}, E_{t-q}^{t-1}, C_t, \theta_Y) \right. \\
\left. + \log p(E_t|\theta_E)\right]
$$

Defining $X_t = (E_{t-q}^{t-1}, Y_{t-p}^{t-1}, C_t)$ and $\phi = (\beta, \alpha, \eta)$ we use the $\sigma\text{ARMA}^{xp}$ model definition (5) to obtain

$$
\begin{aligned}
Q(\theta|\Theta^n) &= \frac{1}{2} \sum_t \mathbb{E}_{\Theta^n}\left[\left(Y_t - (\zeta + \phi X_t^\top + \beta_0 E_t)\right)^2 / \sigma\right] \\
&\quad - \frac{1}{2} \sum_t \mathbb{E}_{\Theta^n}\left[E_t^2 / \gamma\right] + c \quad (7)
\end{aligned}
$$

where $c$ is a constant, $\theta = (\zeta, \phi, \gamma)$ is the free parameters in the model, $\beta_0 = 1$, and $\top$ denotes transpose.

**M-step**

The ML estimation for $\gamma$ is easily obtained as the expected sample variance. That is,

$$\hat{\gamma} = \sum_t \mathbb{E}\left[E_t^2\right] / (T - R) \quad (8)$$

where we have eased the notation by letting $\mathbb{E}$ denote expectation with respect to $\Theta^n$.

Differentiating (7) we obtain the following partial derivatives

$$\begin{aligned}
\frac{dQ}{d\phi} &\propto \sum_t \mathbb{E}\left[Y_t X_t^\top\right] - \sum_t \mathbb{E}\left[X_t^\top X_t\right]\phi^\top \\
&\quad - \sum_t \mathbb{E}\left[X_t^\top E_t\right] - \sum_t \mathbb{E}\left[X_t^\top\right]\zeta \\
\frac{dQ}{d\zeta} &\propto \sum_t \mathbb{E}\left[Y_t\right] - \sum_t \mathbb{E}\left[X_t\right]\phi^\top \\
&\quad - \sum_t \mathbb{E}\left[E_t\right] - \sum_t \zeta \qquad (9)
\end{aligned}$$

By setting derivatives to zero and solving this set of equations we obtain the ML estimate for the parameters $(\zeta, \phi)$. The set of equations can be singular (or close to singular) so one should be careful and use a method robust to this situation when solving the equations. We use pseudo-inversion to extend the notion of inverse matrices to singular matrices when solving the equations (see, e.g., Golub and Van Loan, 1996).

By realizing that the $\sigma\text{ARMA}^{xp}$ model defines a Bayes net, the expected sufficient statistics involved in the above ML estimation can be efficiently computed by the procedure described in Lauritzen and Jensen (2001), or any other stable method for inference in Gaussian Bayes nets. In the use of this procedure, we define an inference structure with cliques $(Y_t, E_t, X_t)$, $t = R+1, \ldots, T$ for each time series in the model. Except for the first clique in a time series, all cliques are initially assigned the densities $p(Y_t, E_t|X_t)$ given by the current parameterization. The first clique is assigned the densities $p(Y_t, E_t, X_t \setminus C_t | C_t)$.

### 3.2 $\sigma\text{ARMA}^{*xp}$ (free $\beta_0$)

The set of free parameters in an $\sigma\text{ARMA}^{*xp}$ model additionally include the parameter $\beta_0$. Therefore this time, let $X_t = (E_{t-q}^t, Y_{t-p}^{t-1}, C_t)$ and $\phi = (\beta_0, \beta, \alpha, \eta)$. Differentiating (7) with respect to this alternative set of free parameters, we obtain the following partial derivatives

$$\begin{aligned}
\frac{dQ}{d\phi} &\propto \sum_t \mathbb{E}\left[Y_t X_t^\top\right] - \sum_t \mathbb{E}\left[X_t^\top X_t\right]\phi^\top \\
&\quad - \sum_t \mathbb{E}\left[X_t^\top\right]\zeta \\
\frac{dQ}{d\zeta} &\propto \sum_t \mathbb{E}\left[Y_t\right] - \sum_t \mathbb{E}\left[X_t\right]\phi^\top - \sum_t \zeta \quad (10)
\end{aligned}$$

Hence, the M-step is performed by solving (8) and solving the slightly simpler set of equations (10) with one more unknown/equation than (9).

### 3.3 EM Only Works for $\sigma > 0$

It turns out that the convergence rate for EM decreases with smaller (fixed) values of $\sigma$. In particular, the EM algorithm will not be able to improve the initial parameter setting in any significant way when $\sigma = 0$. To realize this fact, let $\sigma \to 0$. In this case, $Y_t \to \zeta + X_t\phi' + E_t$ for the $\sigma\text{ARMA}^{xp}$ representation and $Y_t \to \zeta + X_t\phi'$ for the alternative $\sigma\text{ARMA}^{*xp}$ representation. (Recall that $X_t$ and $\phi$ are defined differently for the two representations.) By inserting into (9) and (10), respectively, we see that in both cases

$$\begin{aligned}
\frac{dQ}{d\phi} &\to 0 \\
\frac{dQ}{d\zeta} &\to 0
\end{aligned}$$

Hence, the EM algorithm will not be able to improve the $(\zeta, \phi)$ parameters in any of the model representations. In both representations the $\gamma$ variance parameter will improve at the first EM step, but because the expected statistics used to compute (8) are not a function of this variance, and because the remaining parameters in both model representations do not change, the models will not improve further.

## 4 $\sigma\text{ARMA}^{xp}$ Forecasting

We now consider the problem of using a $\sigma\text{ARMA}^{xp}$ or $\sigma\text{ARMA}^{*xp}$ model to forecast. For a given time series sequence in the model, the task of forecasting is to calculate the distributions for future observations given a previously observed sequence of observations for that time series and sequences of observations for related cross-predictors. We distinguish between two important types of forecasting: (1) *one-step forecasting* and (2) *multi-step forecasting*. In our evaluation of predictive accuracy, we use one-step forecasting.

In one-step forecasting, we are interested in predicting $Y_{T+1}$ given any known observations on the observable variables $Y_1^T$ and $C_1^{T+1}$. For this situation, the predictive posterior distribution can be computed as follows. Using Bayes-net inference we (pre-)compute the marginal posterior distribution $p(E_{T-q+1}^T, Y_{T-p+1}^T | y_1^T, c_1^T)$ for the last clique associated with the time series in the inference structure, as defined in Section 3.1. Recall that some of the observable variables may have missing observations. We first consider the simpler situation where $y_{T-p+1}^T$ and $c_{T+1}$ are completely observed. Let $\mathbb{E}[E_{T-q+1}^T]$ and $\Sigma$ denote the mean vector and covariance matrix for the marginal multivariate Normal distribution $p(E_{T-q+1}^T | y_1^T, c_1^T)$. With $y_{T-p+1}^T$ and $c_{T+1}$ completely observed we can derive the predictive posterior distribution $p(Y_{T+1} | y_1^T, c_1^{T+1})$ as the Normal distribution

$N(\mu^*, \sigma^*)$ with mean and variance

$$\mu^* = \zeta + \mathbb{E}[E_{T-q+1}^T]\beta^\top + y_{T-p-1}^T\alpha^\top + c_{T+1}\eta^\top \quad (11)$$
$$\sigma^* = \sigma + \beta\Sigma\beta^\top + \gamma \quad (12)$$

This derivation can be verified by following the same inference step as for the situation with missing observations, described below.

Notice from (12) that if we compare to an ARMA model, where the variance for the one-step predictive distribution is $\gamma$, a $\sigma\text{ARMA}^{xp}$ model "smoothes" the distribution by adding $\sigma + \beta\Sigma\beta^\top$ to this variance. As $\Sigma$ is positive definite and hence $\beta\Sigma\beta^\top$ is positive, the fixed observation variance $\sigma$—specified by the user—can therefore be conceived as the lowest allowed level for variance in the predictive distribution.

If any of the variables in $Y_{T-p+1}^T$ and $C_{T+1}$ are not observed, it becomes more complicated to compute the posterior predictive distribution for $Y_{T+1}$. In this case, we extend the inference structure with an additional clique $(Y_{T-p+1}^{T+1}, E_{T-p+1}^{T+1}, C_{T+1})$ initially assigned the conditional densities $p(Y_{T+1}, E_{T+1}|Y_{T-p+1}^T, E_{T-p+1}^T, C_{T+1})$ defined by the current parameterization. We then insert any observations we may have on $C_{T+1}$ into this clique and perform a simple Bayes net inference from the (pre-)computed $p(E_{T-q+1}^T, Y_{T-p+1}^T|y_1^T, c_1^T)$ to the new clique by which we obtain $p(E_{T-q+1}^{T+1}, Y_{T-p+1}^{T+1}|y_1^T, c_1^{T+1})$. The posterior predictive distribution is then created by simple marginalization to $Y_{T+1}$. It should be noted that this predictive distribution is only an approximation if some of the variables in $C_{t+1}$ are not observed. We choose this approximation in order to avoid complete inference over all time series in the model and in this way allow for fast predictions.

In multi-step forecasting, we are interested in predicting values for variables at multiple future time steps. When forecasting more than one step into the future we introduce intermediate unobserved prediction variables. For example, say that we are interested in a three-step forecast. In this case we introduce the variables $Y_{T+1}$, $Y_{T+2}$, and $Y_{T+3}$, where only $Y_{T+3}$ is observed. This situation is similar to the incomplete data situation, described above, and the posterior predictive distribution can be obtained in the same way by introducing a clique in the inference structure for each additional time step.

## 5 Evaluation

In this section, we provide an empirical evaluation of our methods. We use two collections of data sets: US-Econo and JPN-Econo. The data sets are compiled by Economagic and can be obtained for a fee from http://www.economagic.com. The US-Econo collection contains monthly data on 50 economic indicators from February 1992 to December 2002, as reported by the U.S. Census Bureau. All the US-Econo data sets are of length 131. The JPN-Econo collection contains monthly data on 53 economic indicators, as reported by the Bank of Japan and the Economic Planning Agency of Japan. The data sets in this collection vary in length from 52 to 259 periods with both the median and mean lengths larger than 192 periods (16 years). The longest time series begin in January 1981 and all end in July 2002.

Each data set is standardized before modeling—that is, for each variable we subtract the mean value and divide by the standard deviation. We divide each data set into a training set, used as input to the learning method, and a holdout set, used to evaluate the models. We use the last twelve observations as the holdout set, knowing that the data are monthly.

In order to evaluate our models on incomplete data, we have created—for each of the two data collections—five corresponding incomplete data collections with respectively 10, 20, 30, 40, and 50 percent of randomly missing observations in each of the training sets in a collection. Corresponding to the incomplete data collections, we have also created corresponding collections of filled-in data sets, where a missing observation is filled in by linear interpolation or extrapolation of the two closest (observed) data points in the training set.

In our experiments, we perform the following comparisons. One, for the complete data collections, predictions for ARMA models learned by the Levenberg-Marquardt method for ML estimation are compared to predictions for the $\sigma\text{ARMA}$ and $\sigma\text{ARMA}^*$ models. This comparison will illustrate the smoothing effect of stochastic ARMA models. Two, to illustrate the importance of handling missing data in a theoretically sound way, we compare predictions for $\sigma\text{ARMA}$ models trained on the incomplete data sets with those for $\sigma\text{ARMA}$ models trained on the filled-in data. Three, the effect of cross predictors is illustrated by comparisons of predictions for $\sigma\text{ARMA}^{xp}$ and $\sigma\text{ARMA}^{*xp}$ models with predictions for $\sigma\text{ARMA}$ and $\sigma\text{ARMA}^*$, respectively, all trained on the complete data sets.

We evaluate the quality of a learned model by computing the *sequential predictive score* for the holdout data set corresponding to the training data from which the model was learned. The sequential predictive score for a model is simply the average log-likelihood obtained by a one-step forecast for each of the observations in the holdout set. To evaluate the quality of a learning method, we compute the average of the sequential predictive scores obtained for each of the time series in a collection. Note that the use of the log-likelihood to measure performance simultaneously evaluates both the accuracy of the estimate and the accuracy of the uncertainty of the estimate. Finally, we use a (one-sided) sign test at significance level 5% to evaluate if one method is significantly better than another. To form the sign test, we count the number of times one method improves the predictive score over the other for each individual time series in a collection. Excluding ties, we seek to reject the hypothesis of equality, where the test statistic for the sign test follows a binomial distribution with probability parameter 0.5.

In all experiments we search for the best structure for models—that is, the best $p$ and $q$ for ARMA, $\sigma$ARMA, and $\sigma$ARMA$^*$ models, and additionally the best set of cross-predictors for each time-series sequence in $\sigma$ARMA$^{xp}$ and $\sigma$ARMA$^{*xp}$ models. For the structural search, we split the training data set into a structural training set and a structural validation set. We use the last twelve observations of the training data as the structural validation set. During structural search, we use the structural training set to estimate parameters and we use the structural validation set to evaluate alternative models by the sequential predictive score. After the model structure is selected, we finally use the full training data set to estimate parameters.

We use a greedy search strategy when deciding which models to evaluate during the structural search. The search ARMA$(p,q)$, $\sigma$ARMA$(p,q)$, and $\sigma$ARMA$^*(p,q)$ models has an outer loop that interatively increases $p$ by one starting from $p=0$. For each $p$, we greedily search for the best $q$ by incrementing or decrementing $q$ from its current value. (We initialize $q$ to zero for $p=0$). The search stops when increasing $p$ and searching for the best $q$ does not improve the model over the model selected for the previous $p$.

The search strategy for $\sigma$ARMA$^{xp}$ and $\sigma$ARMA$^{*xp}$ models is modified to handle cross predictors. In these experiments, all time-series sequences are completely observed. A $\sigma$ARMA$^{xp}$ model therefore factorizes into the individual time series with cross predictors, and each components is learned independently as follows. We first create a ranked list of cross predictors expected to have the most influence on the time-series component. Cross predictors are ranked according to the sequential predictive score on the structural validation data for models with $p=q=0$ and only one cross predictor. We then greedily search for $p$, $q$, and the cross predictors. In the outer loop, we increase $p$ by one starting from $p=0$. In a middle loop, we greedily search for the best $q$ by incrementing or decrementing $q$ from its current value. (Again, we initialize $q$ to zero for $p=0$.) In a inner loop, we add cross predictors in rank order until the score does not increase. If the first added cross predictor does not increase the score, we delete cross predictors in reverse rank order until the score does not increase. (Initially, for $p=q=0$, no cross predictors are included.) The search stops when increasing $p$ and searching for the best $q$ and the best cross-predictors does not improve the model over the model selected for the previous $p$.

Table 1: ARMA, smoothed ARMA, and $\sigma$ARMA average sequential predictive scores for complete data collections.

| COLLECTION | METHOD | AVE. SCORE |
|---|---|---|
| US-ECONO | ARMA | -4.928 |
| US-ECONO | SMOOTHED ARMA | -4.599 |
| US-ECONO | $\sigma$ARMA (FIXED $\beta_0$) | -4.520 |
| US-ECONO | $\sigma$ARMA$^*$ (FREE $\beta_0$) | -4.533 |
| JPN-ECONO | ARMA | -2.935 |
| JPN-ECONO | SMOOTHED ARMA | -2.095 |
| JPN-ECONO | $\sigma$ARMA (FIXED $\beta_0$) | -2.023 |
| JPN-ECONO | $\sigma$ARMA$^*$ (FREE $\beta_0$) | -1.994 |

Recall from Section 4 that the stochastic models proposed in this paper smooth the parameters in the model in a way that sets the lowest allowable variance for the predictive distribution as given by the user specified observation variance, $\sigma$. Similarly, the predictive distribution for an ARMA model can be smoothed by adding $\sigma$ directly to the variance of the predictive distribution. In our first comparison, we show that this *ad hoc* smoothing method is not as good as the smoothing afforded by our stochastic models. We perform these comparisons using $\sigma = 0.01$, although the results are not sensitive to this value.

## 6 Results

We first consider the predictive accuracy of ARMA versus $\sigma$ARMA and $\sigma$ARMA$^*$ models on single time series. It is conventional wisdom that most time-series data can be adequately represented by ARMA$(p,q)$ models with lag $R = max(p,q)$ not greater than two (see, e.g., Box, Jenkins, and Reinsel, 1994). To verify that our model selection method with *unrestricted* lag does not overfit, we ran experiments where we restricted the maximum lag $R$ at varying values. As we increased $R$, the average predictive score for all experiments increased for all three model classes. In fact, the average predictive score for unrestricted experiments was the best but not significantly better than that for any restricted lag. In the following, we will report results obtained for unrestricted lag.

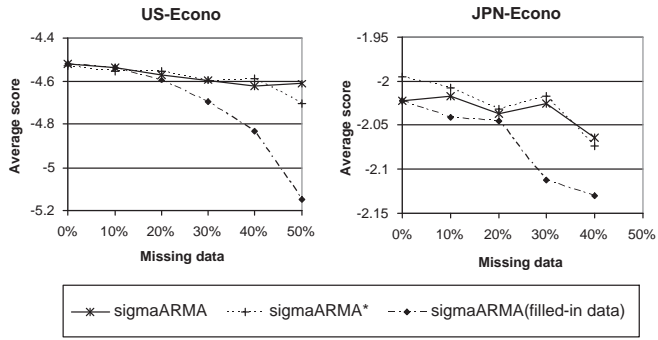

Figure 3: Average sequential predictive scores for complete and incomplete data collections.

Table 2: $\sigma\text{ARMA}^{xp}$ and $\sigma\text{ARMA}^{*xp}$ average sequential predictive scores for complete data collections.

| COLLECTION | METHOD | AVE. SCORE |
|---|---|---|
| US-ECONO | $\sigma\text{ARMA}^{xp}$ (FIXED $\beta_0$) | -4.506 |
| US-ECONO | $\sigma\text{ARMA}^{*xp}$ (FREE $\beta_0$) | -4.526 |
| JPN-ECONO | $\sigma\text{ARMA}^{xp}$ (FIXED $\beta_0$) | -2.019 |
| JPN-ECONO | $\sigma\text{ARMA}^{*xp}$ (FREE $\beta_0$) | -1.980 |

Table 1 shows average sequential predictive scores for each of the two collections of complete data sets—US-Econo and JPN-Econo—for ARMA, smoothed ARMA, $\sigma$ARMA (fixed $\beta_0$), and $\sigma\text{ARMA}^*$ (free $\beta_0$). Figure 3 shows predictive scores as the data becomes more and more incomplete. In these experiments, $\sigma$ARMA models are trained on both incomplete data and filled-in data. Note that we have not included results for JPN-Econo with 50% missing data in the figure; these results follow the same trend as for less missing data but the scores are too small to show in the chosen data range.

For both collections, we see that smoothing matters: $\sigma$ARMA and $\sigma\text{ARMA}^*$ predict more accurately than smoothed ARMA, and smoothed ARMA predicts more accurately than ARMA. All differences are significant by the sign test described previously. In contrast, the predictive accuracy of $\sigma$ARMA and $\sigma\text{ARMA}^*$ are not significantly different. We also see that treating missing data in a principled way is important. In particular, as the data becomes more and more incomplete, $\sigma$ARMA learned using EM with missing data yields more accurate predictions than does $\sigma$ARMA learned using filled-in data. The differences are significant for 30% missing data and more for both data collections.

Now we consider multiple time series and the effect of cross predictors. Table 2 shows average sequential predictive scores for $\sigma\text{ARMA}^{xp}$ and $\sigma\text{ARMA}^{*xp}$ models for the two complete data collections. We see that, in all cases, allowing cross-predictors improves the average score slightly. For the US-Econo data collection, the improvement is significant for $\sigma\text{ARMA}^{xp}$. For the the JPN-Econo data collection, the improvement is significant for both $\sigma\text{ARMA}^{xp}$ and $\sigma\text{ARMA}^{*xp}$.

## 7 Related and Future Work

The most common approach to handling incompleteness of data for ARMA models is to transform the model into a state-space model and perform Kalman filtering—and possibly Kalman smoothing (see, e.g., Jones, 1980). As described in Ghahramani (1998) it is well-known that the Kalman filter for an ARMA model can be represented as forward inference in a dynamic Bayesian network with graphical structure as shown in Figure 4, where $X_t$ is a state vector and $Y_t$ an observation variable. Many different state-space representations are possible for an ARMA model, but these representations are not necessarily intuitive and take some engineering to construct. In contrast, the $\sigma$ARMA representation is straight-forward and the graphical representation clearly describes the model structure and as such, the $\sigma$ARMA representation is easy to extend to more sophisticated models.

In a forthcoming paper we plan to show that one filtering step in the Kalman filter for an ARMA model is equivalent to a one step forward propagation of evidence in the clique representation for the $\sigma$ARMA graphical model and similarly for Kalman smoothing and back-propagation. Therefore, any optimization algorithm used to estimate ARMA models in the state-space representation should be applicable to our $\sigma$ARMA (with $\sigma = 0$) representation as well; and we should get the same result in either representation. Conversely, the EM algorithm should be applicable for the state-space representation of a smoothed ARMA model.

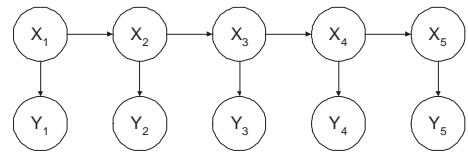

Figure 4: State-space model.

As an alternative to the Kalman-filter approach for handling missing data in ARMA models, Penzer and

Shea (1997) suggest an approach relying on a Cholesky decomposition of the banded covariance matrix for a transformation of the observations in the time series. They demonstrate that this approach is computationally superior for high-order ARMA models with many more AR regressors than MA regressors, and becomes less favorable as the amount of missing data increases, the models become smaller, or more MA regressors are introduced.

Turning our attention to the cross-predictor extension of the $\sigma$ARMA models, the autoregressive tree (ART) models, as defined in Meek *et al.* (2002), allow for selective cross predictors from related time series. ART models, however, are generalizions of AR models and have no MA component. In a slightly different interpretation of cross predictors, Bach and Jordan (2004) are concerned with finding a graphical model structure for entire time series. Their algorithm learns relations between entire time series for spectral representations of the time series.

Finally, as a further generalization of the stochastic ARMA models, defined in this paper, it seems reasonable to let the observation variance $\sigma$ vary freely and estimate this parameter from data as well. However, in preliminary experiments we have experienced that $\sigma$ quickly converges to zero and the EM algorithm will thereafter not be able to improve the remaining parameters (see Section 3.3). We intend to investigate this behaviour further, and hope to find a theoretical explanation.

### Acknowledgments

We thank Jessica Lin for the implementation of ARMA.